# Development of a versatile micro-focused angle-resolved photoemission spectroscopy system with Kirkpatrick-Baez mirror optics

Miho Kitamura,[1] Seigo Souma,[2,3] Asuka Honma,[4] Daisuke Wakabayashi,[1] Hirokazu Tanaka,[1] Akio Toyoshima,[1] Kenta Amemiya,[1] Tappei Kawakami,[4] Katsuaki Sugawara,[2,3,4,5] Kosuke Nakayama,[4,5] Kohei Yoshimatsu,[6] Hiroshi Kumigashira,[1,6] Takafumi Sato,[2,3,4] and Koji Horiba[1,7]*

[1]*Photon Factory, Institute of Materials Structure Science, High Energy Accelerator Research Organization (KEK), Tsukuba 305-0801, Japan*
[2]*Center for Spintronics Research Network, Tohoku University, Sendai 980-8577, Japan*
[3]*Advanced Institute for Materials Research (WPI-AIMR), Tohoku University, Sendai 980-8577, Japan*
[4]*Department of Physics, Graduate School of Science, Tohoku University, Sendai 980-8578, Japan*
[5]*Precursory Research for Embryonic Science and Technology, Japan Science and Technology Agency, Tokyo 102-0076, Japan*
[6]*Institute of Multidisciplinary Research for Advanced Materials (IMRAM), Tohoku University, Sendai 980-8577, Japan*
[7]*Institute for Advanced Synchrotron Light Source, National Institutes for Quantum Science and Technology (QST), Sendai 980-8579, Japan*

\* Author to whom correspondence should be addressed.   Email: horiba.koji@qst.go.jp

**ABSTRACT**

Angle-resolved photoemission spectroscopy using a micro-focused beam spot (micro-ARPES) is becoming a powerful tool to elucidate key electronic states of exotic quantum materials. We have developed a versatile micro-ARPES system based on synchrotron radiation beam focused with a Kirkpatrick-Baez mirror optics. The mirrors are monolithically installed on a stage, which is driven with five-axes motion, and are vibrationally separated from the ARPES measurement system. Spatial mapping of the Au photolithography pattern on Si signifies the beam spot size of 10 μm (horizontal)×12 μm (vertical) at the sample position, which is well suited to resolve the fine structure in local electronic states. Utilization of the micro beam and the high precision sample motion system enables the accurate spatially resolved band-structure mapping, as demonstrated by the observation of a small band anomaly associated with tiny sample bending near the edge of a cleaved topological insulator single crystal.

**I. INTRODUCTION**

Angle-resolved photoemission spectroscopy (ARPES) is a powerful experimental technique to clarify the electronic structure of materials. By utilizing its unique capability that resolves momentum ($k$) of electrons besides energy, ARPES can directly determine the band structure and the Fermi surface, together with quasiparticle dynamics, that govern fundamental physical properties [1-3]. This makes the ARPES technique one of leading spectroscopic tools to study the key signatures of quantum materials, as highlighted by the recent intensive ARPES studies on high-temperature superconductors, topological insulators, and atomic-layer materials [3-7].

The rapid progress in ARPES largely owes to the recent drastic advancement in the instrumentation of ARPES apparatus, such as the improvement of energy and $k$ resolution associated with the development of an electron analyzer that enables two-dimensional (2D) detection of photoelectrons, as well as the utilization of highly brilliant photon sources such as next-generation synchrotron radiation and a vacuum ultraviolet laser. Recently, an extension of the ARPES technique to access additional electron degrees of freedom, namely, spin and time, is also serving as a promising route to further advance ARPES, as exemplified by the spin-resolved ARPES studies of spin-polarized topological and Rashba surface states [8-11] and time-resolved ARPES studies on the photohole decay dynamics of layered materials [12, 13].

Besides time- and spin-resolved ARPES, the spatially resolved ARPES is recently attracting a great deal of attention [14,15]. This is due to the increasing demand to resolve local band structures in exotic quantum materials, as highlighted by the helical edge states of quantum spin Hall insulators (2D topological insulators) [16] and hinge states of higher order topological insulators [17]. Visualization of key electronic states associated with the superconductivity and the quantum phase transition in the heterostructures of atomic layered materials (e.g. twisted bilayer graphene [18]) as well as the Operando analysis of related devices requires an access to the local electronic states in the spatial region of typically a few tens of microns [19-21]. Also, spatially resolved ARPES is useful to study inhomogeneous samples, hard-to-cleave samples, tiny single crystals, and the samples that contain multiple domains (ferromagnetic, ferroelectric, crystal domains etc).

A straightforward way to achieve high spatial resolution in ARPES is to focus the photon beam at the sample position down to the scale of interest (note that the small beam spot is required from the spectroscopic difficulty in simultaneously imaging the angular and spatial distributions of photoelectrons [22]). In fact, some ARPES apparatus with the spot size from 50 nm to a few 100 nm (nano-ARPES) equipped with the focusing optics of a Fresnel zone plate, a Schwarzschild objective, or a micro capillary are currently in operation in third-generation synchrotron facilities [23-27], whereas these systems may suffer from an experimental difficulty in obtaining a high photoelectron count rate and also in varying photon energy ($h\nu$) while always keeping the focused beam on the sample. On the other hand, there is a huge demand of high-throughput ARPES measurements with high statistics band-structure mapping that do not require an ultimate spatial resolution, as can be recognized from highly productive micro-ARPES beamlines currently in operation in a few synchrotron facilities [25, 28-30]. Such micro-ARPES systems achieve the micro-focused beam by cutting the high-flux photons produced from an undulator by an aperture slit and makes the slit point as the origin of photon emission. However, in most of the micro-ARPES beamlines, the focusing distance in the sample side becomes rather long because the focusing mirror is placed outside the ARPES measurement chamber; this leads to the beam spot size of typically $50 \times 50$ $\mu m^2$ [28-31]. Thus, the development of micro-ARPES system with higher spatial resolution is desired.

In this paper, we report a new micro-ARPES system developed at the undulator beamline BL-28A of the Photon Factory, KEK. By placing a Kirkpatrick-Baez (K-B) mirror optics next to (but outside) the ARPES measurement chamber while keeping the short sample-mirror distance of 400 mm, we have achieved the micro-beam spot of $10 \times 12$ $\mu m^2$ at the sample position. We demonstrate that the obtained micro beam combined with high precision sample motion system is highly useful to detect tiny local modulations in the band structure, by showing spatially modulated band structures of a prototypical topological insulator $Bi_2Se_3$.

**II. OPTICS DESIGN**

First, we explain the design of micro-focusing optics for the micro-ARPES measurements. Figure 1 shows a schematic of the BL-28A beamline, which has been upgraded with K-B mirror optics. A variable-angle Monk-Gillieson-type varied-line-spacing plane-grating monochromator which has no entrance slit [32] is adopted. The light emitted from the undulator is focused both horizontally and vertically by an M0 toroidal mirror located at 15.5 m from the light source. The horizontal focusing point is 23.4 m from the source and the reduction ratio is 15.5 : 7.9 ~ 2 : 1. In the vertical direction, the light is focused by the M0 toroidal mirror and the grating onto the exit slit at 32.0 m from the source. The exit slit plays a role of resolving the photon energy in the monochromator and also serves as a virtual light source in the vertical direction for the post-focusing mirror. The K-B mirror optics and an ARPES measurement chamber are installed about 4 m behind the exit slit. The K-B mirror optics consists of two elliptical mirrors for focusing the beam from horizontal virtual light source (horizontal slit) and the vertical virtual light source (exit slit) independently to the measurement sample.

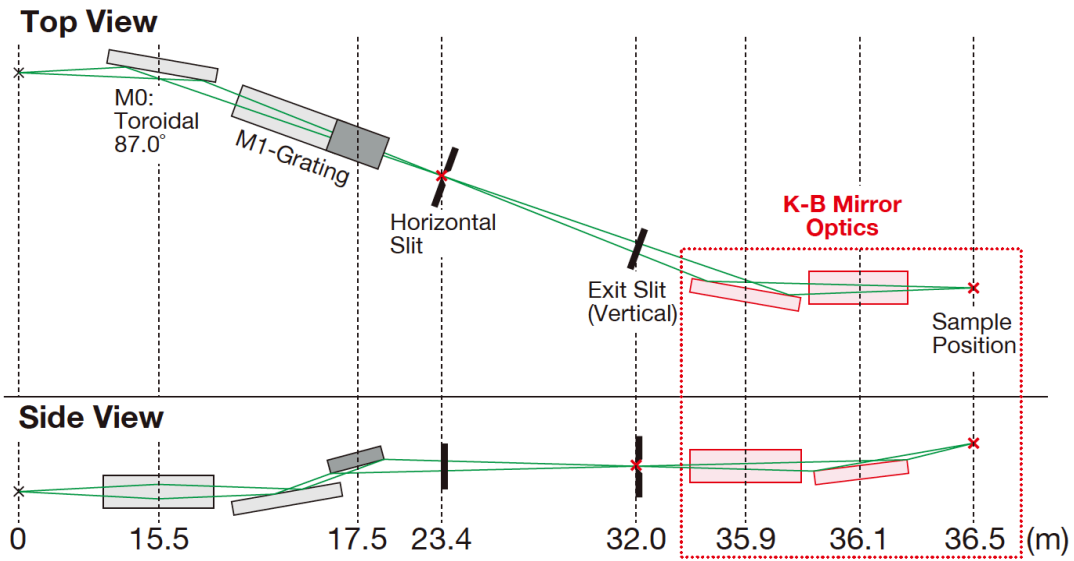

**FIG. 1.** A schematic of the new layout of beamline BL-28A at the Photon Factory with updated K-B mirror optics.

Figure 2(a) shows a schematic diagram of the K-B mirror optics in an actual scale. The first elliptical mirror, which focuses the beam horizontally, has a focal length of 600 mm. The incident angle is set at 3 degrees in order to catch fully the spread light to the horizontal direction without any loss of the beam. The second elliptical mirror, which focuses the beam vertically, has a focal length of 400 mm and an incidence angle of 1 degree to pass through the horizontally-aligned beam port of the existing ARPES system. Both mirrors are fabricated with very high precision, within 1 μrad root mean square (RMS) of the slope error and 0.2 nmRMS of the surface roughness. Figure 2(b) shows the 3D drawing of the K-B mirror system including the K-B mirrors inside a chamber and motorized stages. The two mirrors are monolithically installed on a stage, vibrationally separated from the chamber. The monolithic mirror stage has five-axes motion: translational drive in the *xyz* direction and pitch-rotation drive for both the first mirror ($\theta z$) and the second mirror ($\theta x$). The effect of the yawing rotation of the other mirror due to the pitch rotation of each mirror can be ignored because the adjustment angle is comparatively small.

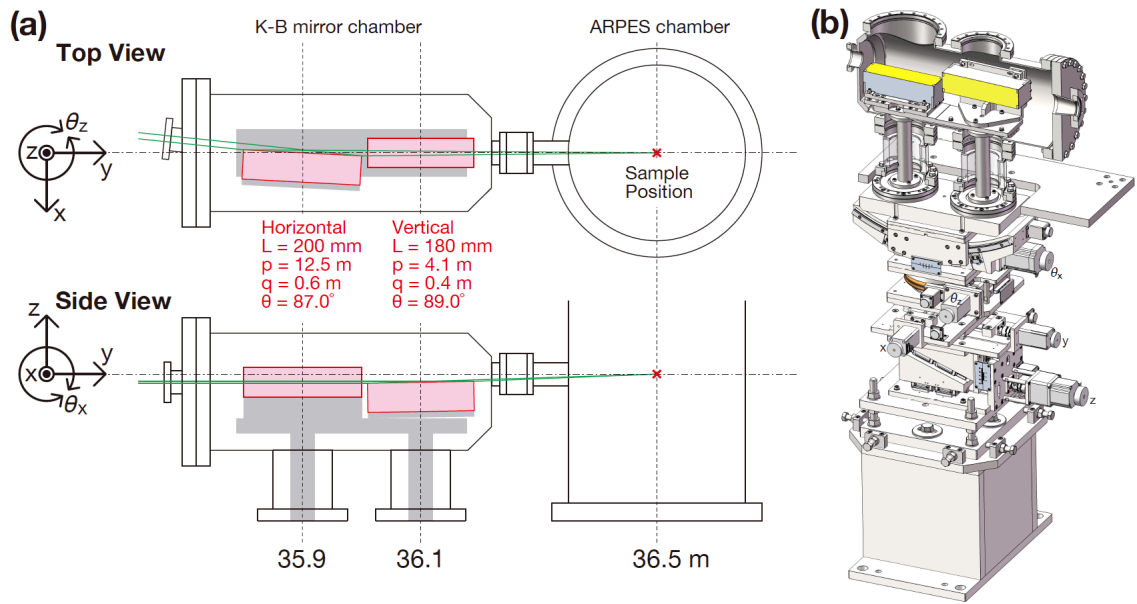

**FIG. 2.** (a) The schematic diagram of K-B mirror optics in an actual scale. The parameters L, p, q, and θ represent the mirror length, distance from the virtual source point to the mirror center, distance from the mirror center to the focal point (sample position), the incident angle, respectively. (b) A 3D drawing of the K-B mirror system

To estimate the actual spot size at the sample prior to the construction of the micro-ARPES system, we have performed ray-tracing simulations. The simulations were performed using ShadowOui [33] in the OASYS environment [34]. Note that all sizes described in the following are expressed in full width at half maximum (FWHM). The light source was simulated using the parameters of an elliptically polarized undulator with six variable rows of magnetic arrays [35] used in BL-28 with the period length $\lambda_u$ of 160 mm and the number of periods $n_u$ of 22 [36]. The simulated source size at the photon energy of 100 eV is $x = 1200$ μm and $z = 120$ μm, and the divergence at the source point is $\sigma x = 290$ μrad and $\sigma z = 26$ μrad. The spot size at the horizontal slit focused by M0 mirror as the virtual light source in the horizontal direction is $x = 610$ μm and $z = 570$ μm. Since the reduction ratio of the horizontal focusing by the K-B mirror is 12.5 : 0.6 ~ 20 : 1, it is necessary to limit the virtual source size to 200 μm by the horizontal slit in order to achieve a 10 μm spot on the sample position. This reduces the light intensity to 29%. Concerning the vertical direction, because the vertical reduction ratio is 4.1 : 0.4 ~ 10 : 1, it is necessary to reduce the virtual source size to 100 μm by the exit slit for achieving a 10 μm spot on the sample position.

Figure 3 shows the simulation results of the beam spot on the sample position focused by the K-B mirror. Figure 3(a) shows the results in the case when the horizontal slit is fully open and the exit slit is set to 200 μm. Note that since the exit slit is the energy slit of the monochromator, the slit width affects not only the spot size but also the energy resolution. The spot profile along the horizontal direction is Gaussian shaped with the width of 28 μm, while the profile along the vertical direction is rectangular with the width of 19 μm. The reduction in light intensity due to the dropout from the K-B mirror is 4-7%, indicating that almost all the light rays are received by the mirror. Therefore, the energy resolution and photon flux after passing the K-B mirror optics are considered to be almost equivalent to those of BL-28A before the introduction of the K-B mirror optics: the photon flux is $10^{11}$–$10^{12}$ photons/s and the energy resolution ($E/\Delta E$) is 2x10$^4$ at the photon energy of 65 eV. [37]. Figure 3(b) shows the results when the horizontal slit is 200 μm and the exit slit is 100 μm, achieving the spot size of 10 μm for both horizontal and vertical directions as designed. The light intensity in case (b) is estimated to be around 15 % compared to that in case (a).

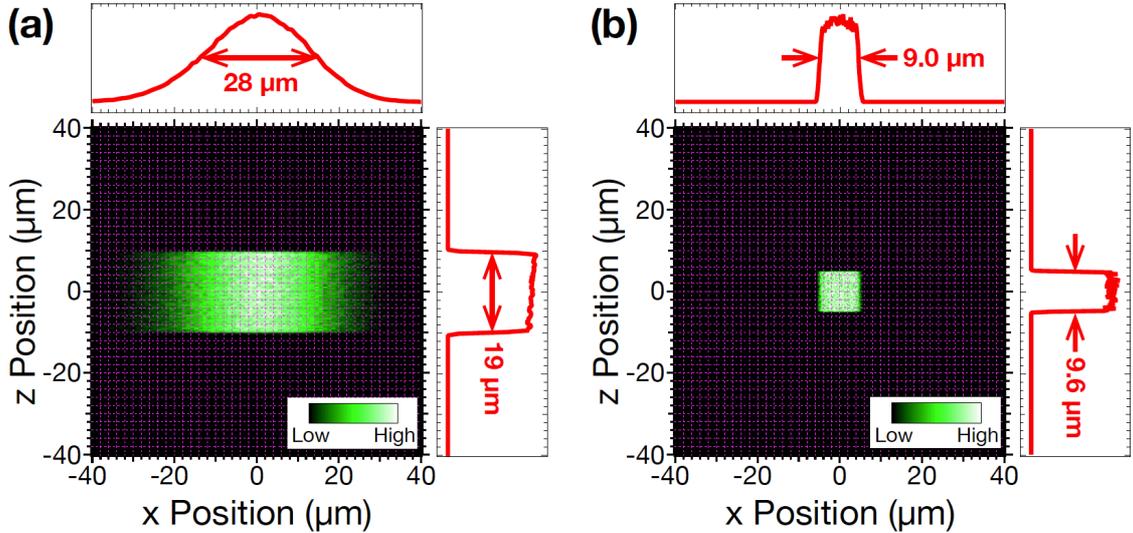

**FIG. 3.** Ray-tracing-simulation results of the beam spot on the sample position. (a) In case that the horizontal slit is fully open and the exit slit is set to 200 μm. (b) In case that the horizontal slit is set to 200 μm and the exit slit is set to 100 μm.

## III. CONSTRUCTION OF MICRO-ARPES SYTEM

Based on the result of the ray-tracing simulation, we have constructed the whole micro-

ARPES system. As shown in Fig. 4(a), the micro-ARPES system mainly consists of (i) the K-B mirror chamber with five-axes motion, (ii) the ARPES measurement chamber (main chamber, originally equipped at the beamline without attaching the K-B mirror), (iii) the electron analyzer, (iv) the sample manipulator with five-axes sample motion (R-dec i-Gonio), and (v) the vacuum pumping system to maintain the whole system in ultrahigh vacuum (UHV). To achieve short focus length, we set the distance between the second mirror and the sample to be 400 mm, by placing the K-B mirror chamber just next to the main chamber (note that we avoided situating the K-B mirror inside the main chamber because it can reduce the flexibility of sample manipulation in the main chamber). The electron analyzer (Omicron-Scienta DA30) has a deflector electron lens system so that the ARPES mapping in 2D $k$ space can be performed without changing the geometrical arrangement between the sample and the synchrotron radiation beam. This is particularly useful when one needs to keep exactly the same beam-spot position on the sample surface during the 2D mapping, which is often the case for tiny single crystals, inhomogeneous samples, and the samples containing multiple domains. The five-axis sample manipulator equipped with a liquid-He cryostat can be traveled in $x$, $y$, and $z$ directions (note that $z$ axis is vertical to the ground) with range of 25, 25, and 300 mm respectively. Polar rotation of +/- 180 degree around the vertical axis is provided by a differential pumping rotary seal. Two sample stages are equipped at the bottom of the manipulator, and thermal radiation to the sample stages is screened by a copper plate (coated by gold) cooled by recycling He gas. One stage has no rotation/tilting capability, directly connected to the cold head of the cryostat, and can be cooled down to 5 K. The other stage is rotatable around tilting axis of +/- 45 degree with respect to the analyzer normal, and can be cooled down to 7 K by utilizing copper braid assembled to the first stage. Sample temperature is controlled by resistive heaters equipped on each stage, and can be increased up to 450 K. The micro-ARPES system is already in operation and open to the users. An actual photograph of the system is shown in Fig. 4(b).

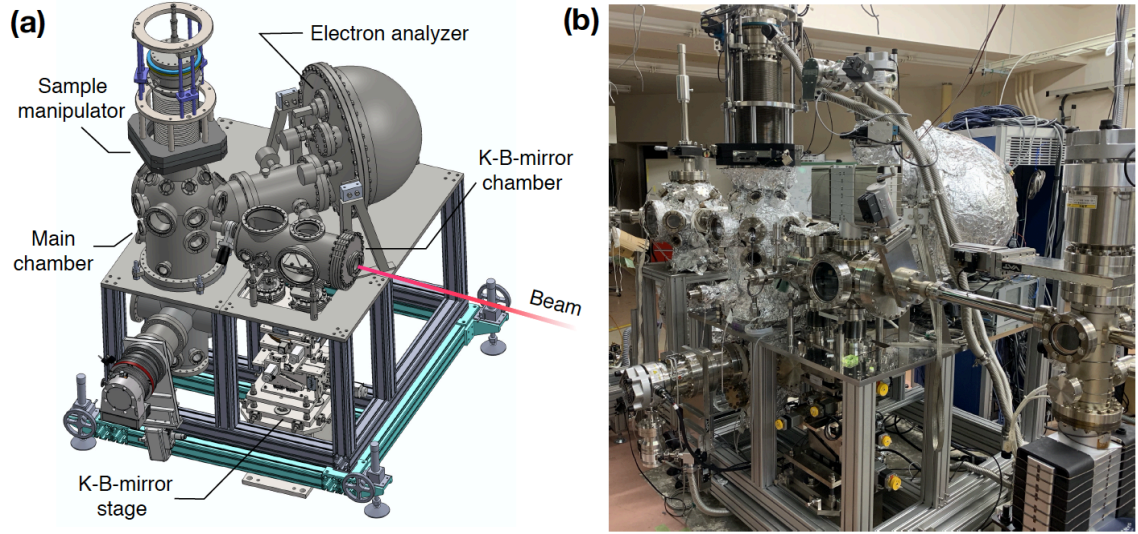

**FIG. 4.** (a) A 3D schematic drawing and (b) a photograph of the micro-ARPES system.

### IV. PERFORMANCE OF MICRO-ARPES SYSTEM

To achieve an accurate micro-ARPES measurement, it is essential to control the sample position with high precision and high reproducibility in addition to the micro-focused synchrotron beam. For this sake, we attached absolute-type optical encoders on all the $xyz$ stages of the sample manipulator (outside the UHV) and monitored each reading ($x$, $y$, $z$). Figure 5(a) shows the reading of ($x$, $y$, $z$) under the stable condition without the sample movement by stepping motors. We found that the readings fluctuate for all the axes without any detectable drift in a time period of the measurement. Such fluctuations originate from low-frequency environmental vibrations with the frequencies of typically below several tens of Hz, originating from machines, moving objects and walking humans etc [38]. We found that the amplitude of fluctuations for the $x$- or $y$-axis is twice as large as that for the *z*-axis (a gravity direction). This is likely because the manipulator suppresses more strongly the vibration along the *z*-axis due to its heavy weight. Such a difference is better visualized from the histogram of the sample position in Fig. 5(b), from which we can estimate the typical vibration width ($\Delta x$, $\Delta y$, $\Delta z$) to be ~200 nm for $\Delta x$ and $\Delta y$, and ~80 nm for $\Delta z$. Since these fluctuations are measured at the stage of the sample manipulator outside the UHV, it is inferred that the actual vibration width on the exact sample position is larger than the readings. However, it would be still far smaller than the beam spot size. Therefore, it is expected that the sample vibration plays a little role to the spatial resolution of the micro-APRES system.

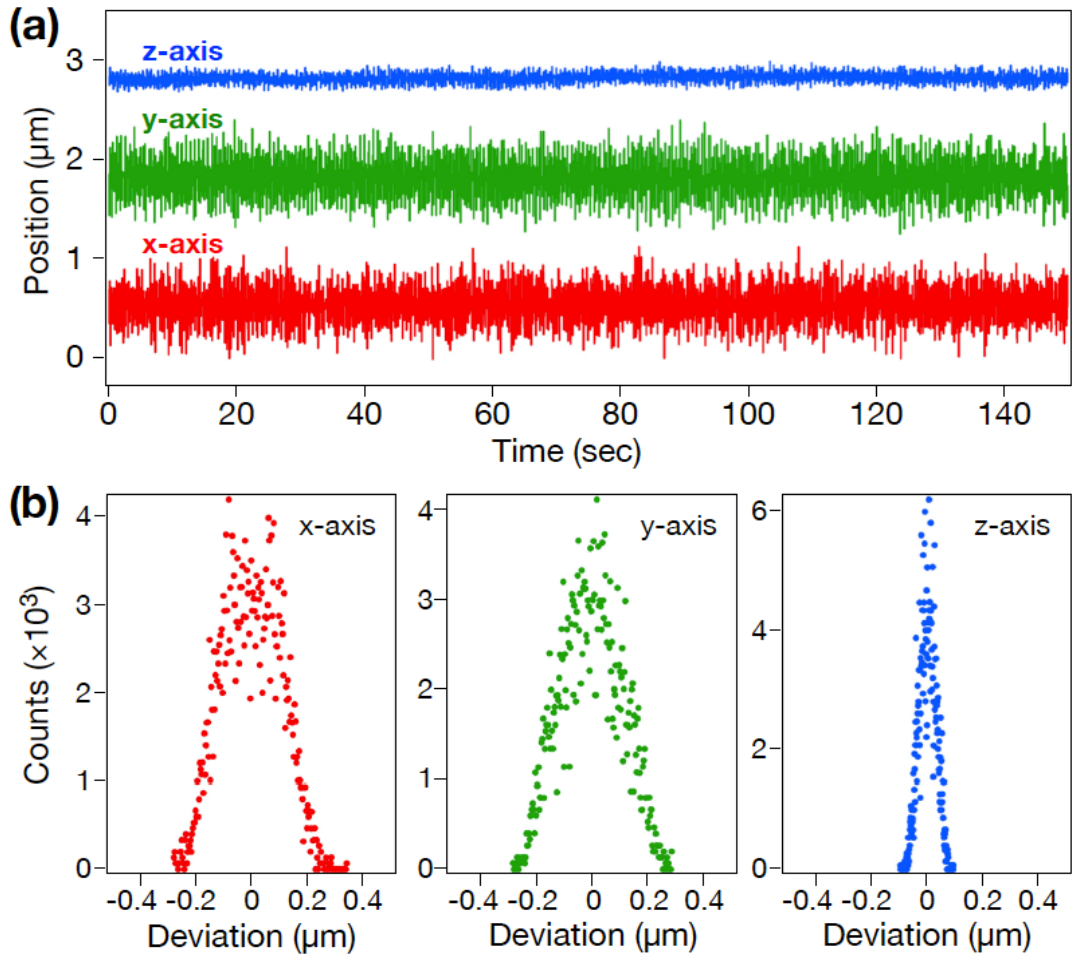

**FIG. 5**. (a) Readings of the optical encoders along *x*, *y*, and *z* axes of the sample manipulator plotted against measurement time. (b) Histograms of (a) for *x*, *y*, and *z*-axes.

The spot size at the actual measurement condition in the micro-ARPES system was evaluated by measuring the Au 4*f* photoemission spectra of Au (50 nm) / Ti (50 nm) pattern samples fabricated by photolithography on Si substrates. Figures 6(a) and 6(b) show the total spectral weight of the Au $4f_{7/2}$ peak plotted against the position, which were obtained by line scans across the edge of the patterns along the horizontal and vertical directions, respectively. The energy of the incident light was 200 eV, and the aperture size of the exit slit was set to 100 μm and that of the horizontal one to 200 μm. The spot size on the sample was estimated to be 14 μm and 12 μm in the horizontal and vertical directions, respectively, by FWHM of the first derivative of the edge profiles [Figs. 6(c) and 6(d)]. Considering that the incident angle of the light was tilted by 45 degrees from the normal to the sample surface in the horizontal direction, the focused beam size was achieved to be 10 μm (horizontal) × 12 μm (vertical). Using this micro-focused beam, we have demonstrated the photoemission measurements on a sample with the letter pattern "ARPES". The mapping result of the Au $4f_{7/2}$ spectral weight is shown in Fig. 6(e). The pattern of the letter "ARPES" was clearly observed. The thickness of the letter lines is around 30 μm, indicating that the photoemission spectral mapping with a spatial resolution of the order of 10 μm can be obtained using this micro-ARPES system with the K-B mirror optics. At our micro-ARPES system, the energy resolution was typically set to 15 meV for the incident light of 90 eV.

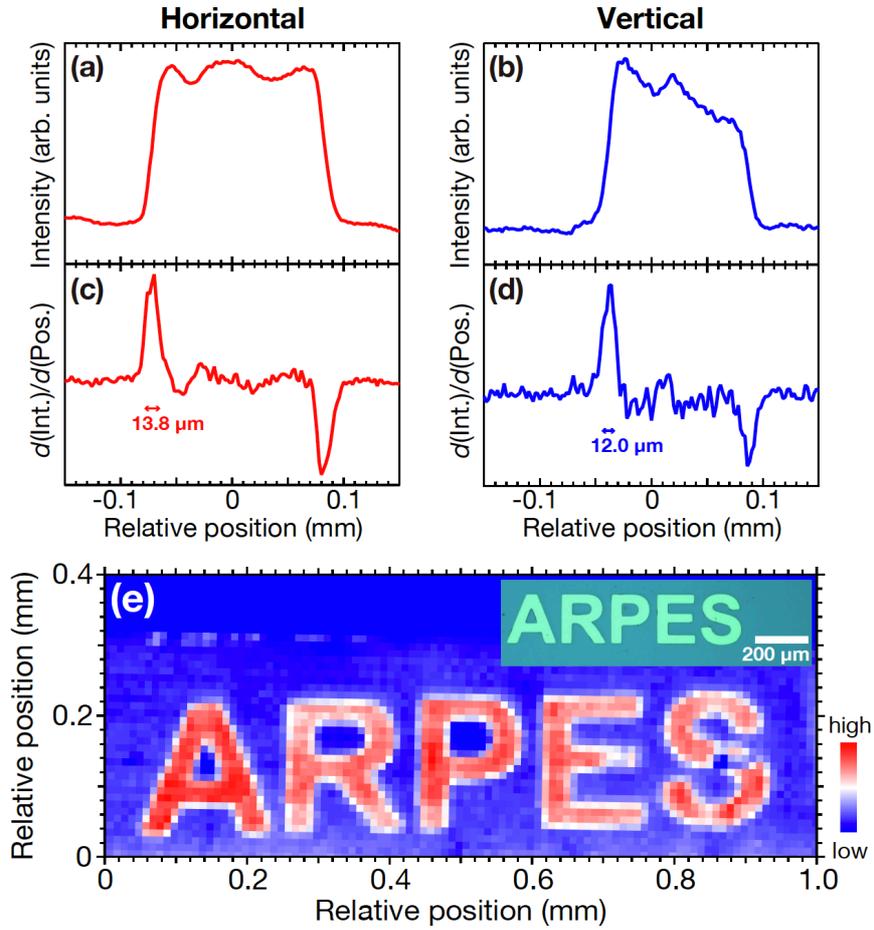

**FIG. 6.** (a) Line profiles of the spectral weight of Au $4f_{7/2}$ peak across the edge of the patterns along the horizontal direction and (b) along the vertical direction. (c) The first derivative of the line profile along the horizontal direction shown in (a). (d) The first derivative of the line profile along the vertical direction shown in (b). (e) The mapping results of the Au $4f_{7/2}$ peak area of the test sample with the letter pattern "ARPES". The inset shows the optical microscope image of the test sample.

Photon-energy dependence of the spot size has been evaluated using a cleaved surface of a $Bi_2Se_3$ crystal, where the spatial distribution of the electronic structure is discussed in detail later. Figure 7(a) shows the spatial mapping of the Bi $4f$ core level peak with the binding energy of around 25 eV taken at the photon energy of 200 eV. Using the selected crystal edges of this sample, we have performed the line scans along the *x* and *z* directions at the two different incident photon energies of 200 eV and 60 eV. The spot sizes estimated by FWHM of the first derivative of the edge profiles are almost the same between the photon energies of 200 eV and 60 eV, as shown in Figs. 7(b) and 7(c). This result indicates that our K-B mirror optics achieves micro-focusing without chromatic aberration even at the low excitation photon energies used for ARPES measurements. When the horizontal slit is restricted to 200 μm to focus the spot size of horizontal direction to ~10 μm, the photon flux reduces to 47% at 200 eV and 36% at 60 eV. Both the values are better than those predicted by the ray-tracing simulations.

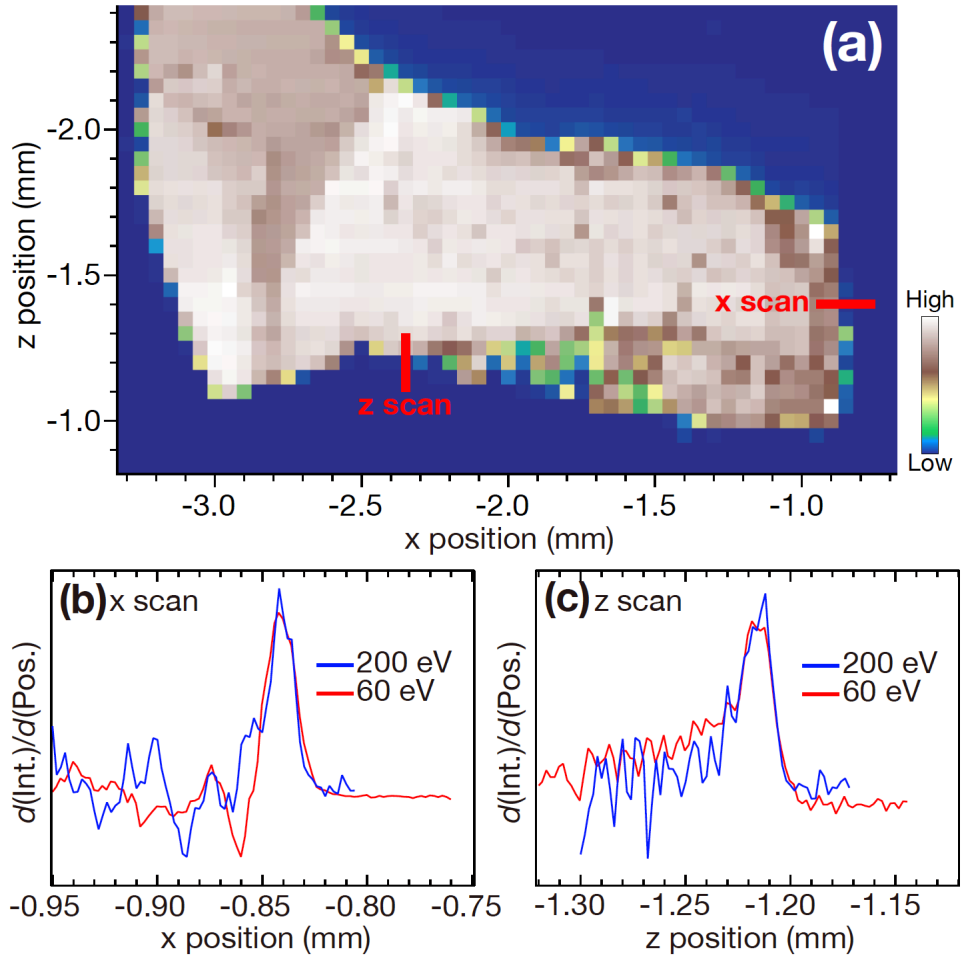

**FIG. 7.** (a) The spatial mapping of the Bi 4*f* peak area on the cleaved $Bi_2Se_3$ bulk crystal,

obtained at the photon energy of 200 eV. Red lines represent real-space position ($x$ scan and $z$ scan) where the line profile shown in (b) and (c) were obtained. (b), (c) The first derivative of the line profile along the $x$- and $z$-scan lines, respectively.

## V. EXAMPLE OF MICRO-ARPES MEASUREMENT

To demonstrate the performance of the micro-ARPES system, we have carried out spatially resolved ARPES measurements on a cleaved surface of a bulk crystal of the prototypical topological insulator $Bi_2Se_3$. $Bi_2Se_3$ is known to have simple band structures characterized by bulk valence and conduction bands at the $\bar{\Gamma}$ point of the surface Brillouin zone, together with the Dirac-cone surface state traversing these bulk bands [39, 40], which is thus particularly suited for the ARPES measurements, as established by accumulated previous ARPES studies on this material [4, 5]. Since each $Bi_2Se_3$ layer in the crystal is weakly coupled with van der Waals force, it is thought to be rather easy to obtain a flat surface after cleaving the $Bi_2Se_3$ crystal. In fact, an optical microscope image shown in Fig. 8(a) displays a shiny flat surface after cleaving, which is also corroborated by the overall uniform spatial image of the ARPES intensity in Fig. 8(b) obtained in the spatial area enclosed by a red rectangle in Fig. 8(a). An ARPES image obtained on the representative spatial point of the flat surface [point A in Fig. 8(b)] signifies bulk conduction and valence bands [Fig. 8(c)], together with a weaker topological surface state (note that the intensity of the surface state is relatively suppressed in this photon energy due to the matrix-element effect), which is a typical signature of the band structure of $Bi_2Se_3$. On the other hand, when the micro-focused beam is irradiated around the edge of the sample (points B-D; within 100 μm from the edge), the conduction band and surface state are not clearly resolved. Interestingly, although top of the valence band, which is located at the angle of $\theta = 0°$ at point A, systematically moves on the right hand direction (positive $\theta$) on moving from point B to point D. Such behavior is satisfactorily explained in terms of the small (about a few degrees) bending of the crystal around the edge of the cleaved sample which is hard to be resolved with the standard ARPES apparatus with the larger beam spot. This result highlights the usefulness of the micro-ARPES system to investigate the spatially resolved electronic states in great details.

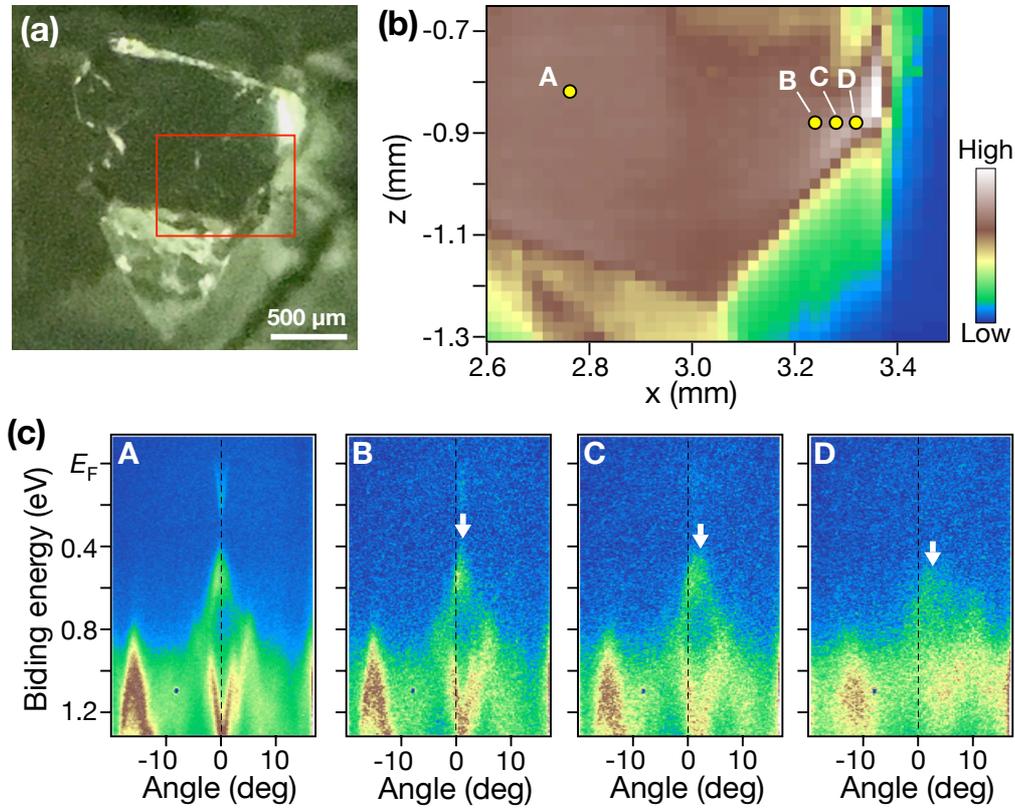

FIG. 8. (a) An optical microscope image for the cleaved $Bi_2Se_3$ single crystal in which the micro-ARPES measurements were performed. (b) A spatial mapping of the ARPES intensity on the cleaved surface of $Bi_2Se_3$ measured at room temperature with $h\nu = 80$ eV. The energy resolution was set as 40 meV. ARPES intensity reflects total intensity integrated over energy and angular axes obtained with the angular mode of the electron analyzer. (c) ARPES intensity as a function of the angle and the binding energy, obtained

at four representative points on the surface (points A-D).

## VI. CONCLUSION

We have developed a versatile micro-ARPES system with the K-B mirror optics. We have achieved a micro beam spot of 10 μm (horizontal) × 12 μm (vertical) at the sample position. Such a small beam spot was achieved by the compact design of the K-B mirror chamber which is placed just next to the ARPES measurement chamber, thereby keeping the sample-mirror distance short. As demonstrated by the observation of local modulation of the electronic band structure around the edge of a topological insulator $Bi_2Se_3$, this micro-ARPES system has a high capability to investigate spatially modulated electronic states of various materials.

**ACKNOWLEDGMENTS**

This work was supported by JST-CREST (No: JPMJCR18T1), JST-PRESTO (No: JPMJPR20A8), JSPS (JSPS KAKENHI No: JP17H01139, JP18K14130, JP26287071, JP20K20906, JP20H02853, JP20H01847, JP21H04435, JP21H01757, JP21K14541), and KEK-PF (Proposal number: 2018S2-001, 2021S2-001). T. Kawakami thanks GP-Spin for financial support. This work is supported by Mechanical Engineering Center, Applied Research Laboratory, KEK.

**DATA AVAILABILITY**

The data that support the findings of this study are available from the corresponding authors upon reasonable request.